\documentclass[fleqn,twoside,xspace]{article}
\usepackage{espcrc2}

\usepackage{graphicx}
\usepackage[figuresright]{rotating}

\RequirePackage{xspace}

\def\lhcb {LHC{\em b\/}\xspace}

\def\btev {BTe\kern -0.1em V}

\def\babar{\mbox{\slshape B\kern-0.1em{\small A}\kern-0.1em B\kern-0.1em{\small A\kern-0.2em R}}}

\def\ee         {\ensuremath{e^-e^-}\xspace}

\def\mup        {\ensuremath{\mu^+}\xspace}
\def\mun        {\ensuremath{\mu^-}\xspace} % muon negative (\mum is taken)

\def\ellell     {\ensuremath{\ell^+ \ell^-}\xspace}

\def\pip   {\ensuremath{\pi^+}\xspace}

\def\Kbar  {\kern 0.2em\overline{\kern -0.2em K}{}\xspace}

\def\Kz    {\ensuremath{K^0}\xspace}
\def\Kzb   {\ensuremath{\Kbar^0}\xspace}
\def\KzKzb {\ensuremath{\Kz \kern -0.16em \Kzb}\xspace}
\def\Kp    {\ensuremath{K^+}\xspace}
\def\Km    {\ensuremath{K^-}\xspace}

\def\KpKm  {\ensuremath{\Kp \kern -0.16em \Km}\xspace}

\def\Kstarzb {\ensuremath{\Kbar^{*0}}\xspace}

\def\Dbar    {\kern 0.2em\overline{\kern -0.2em D}{}\xspace}

\def\Dz      {\ensuremath{D^0}\xspace}
\def\Dzb     {\ensuremath{\Dbar^0}\xspace}
\def\DzDzb   {\ensuremath{\Dz {\kern -0.16em \Dzb}}\xspace}
\def\Dp      {\ensuremath{D^+}\xspace}
\def\Dm      {\ensuremath{D^-}\xspace}

\def\DpDm    {\ensuremath{\Dp {\kern -0.16em \Dm}}\xspace}

\def\Bbar    {\kern 0.36em\overline{\kern -0.18em B}{}\xspace}

\def\Bz      {\ensuremath{B^0}\xspace}
\def\Bzb     {\ensuremath{\Bbar^0}\xspace}
\def\BzBzb   {\ensuremath{\Bz {\kern -0.16em \Bzb}}\xspace}
\def\Bu      {\ensuremath{B^+}\xspace}
\def\Bub     {\ensuremath{B^-}\xspace}

\def\BpBm    {\ensuremath{\Bu {\kern -0.16em \Bub}}\xspace}

\def\Bdb     {\ensuremath{\Bbar_d}\xspace}

\mathchardef\Upsilon="7107
\def\Y#1S{\ensuremath{\Upsilon{(#1S)}}\xspace}% no space before {...}!

\mathchardef\Deltares="7101
\mathchardef\Xi="7104
\mathchardef\Lambda="7103
\mathchardef\Sigma="7106
\mathchardef\Omega="710A

\def\Deltabar{\kern 0.25em\overline{\kern -0.25em \Deltares}{}\xspace}
\def\Lbar{\kern 0.2em\overline{\kern -0.2em\Lambda\kern 0.05em}\kern-0.05em{}\xspace}
\def\Sigbar{\kern 0.2em\overline{\kern -0.2em \Sigma}{}\xspace}
\def\Xibar{\kern 0.2em\overline{\kern -0.2em \Xi}{}\xspace}
\def\Obar{\kern 0.2em\overline{\kern -0.2em \Omega}{}\xspace}
\def\Nbar{\kern 0.2em\overline{\kern -0.2em N}{}\xspace}
\def\Xb{\kern 0.2em\overline{\kern -0.2em X}{}\xspace}

\newcommand{\tev}{\ensuremath{\mathrm{\,Te\kern -0.1em V}}\xspace}
\newcommand{\gev}{\ensuremath{\mathrm{\,Ge\kern -0.1em V}}\xspace}
\newcommand{\mev}{\ensuremath{\mathrm{\,Me\kern -0.1em V}}\xspace}
\newcommand{\kev}{\ensuremath{\mathrm{\,ke\kern -0.1em V}}\xspace}
\newcommand{\ev}{\ensuremath{\mathrm{\,e\kern -0.1em V}}\xspace}
\newcommand{\gevc}{\ensuremath{{\mathrm{\,Ge\kern -0.1em V\!/}c}}\xspace}
\newcommand{\mevc}{\ensuremath{{\mathrm{\,Me\kern -0.1em V\!/}c}}\xspace}
\newcommand{\gevcc}{\ensuremath{{\mathrm{\,Ge\kern -0.1em V\!/}c^2}}\xspace}
\newcommand{\gevgev}{\ensuremath{{\mathrm{\,Ge\kern -0.1em V^2}}}\xspace}
\newcommand{\mevcc}{\ensuremath{{\mathrm{\,Me\kern -0.1em V\!/}c^2}}\xspace}
\def\invfb   {\ensuremath{\mbox{\,fb}^{-1}}\xspace}

\def\mus  {\ensuremath{\rm \,\mus}\xspace}

\def\mus        {\ensuremath{\,\mu{\rm s}}\xspace}    %% microsecond
\def\to                 {\ensuremath{\rightarrow}\xspace}

\def\pep2{PEP-II}

\def\gsim{{~\raise.15em\hbox{$>$}\kern-.85em
          \lower.35em\hbox{$\sim$}~}\xspace}
\def\lsim{{~\raise.15em\hbox{$<$}\kern-.85em
          \lower.35em\hbox{$\sim$}~}\xspace}

\def\qsq                {\ensuremath{q^2}\xspace}

\def\CP                {\ensuremath{C\!P}\xspace}
\xspace

\newcommand{\app}       [1]  {{Acta Phys.\ Polon.\ {\bf #1}}}

\def\BdbKsmm{\ensuremath{\Bdb \to \Kstarzb\mup\mun}\xspace}
\def\BdbKsll{\ensuremath{\Bdb \to \Kstarzb\ellell}\xspace}

\def\AFB{\ensuremath{A_{\mathrm{FB}}}\xspace}
\def\AT#1{\ensuremath{A_T^{#1}}\xspace}
\def\C#1{\ensuremath{\mathcal{C}_{#1}}}
\def\Cp#1{\ensuremath{\mathcal{C}_{#1}^{'}}}
\def\Ceff#1{\ensuremath{\mathcal{C}_{#1}^{\mathrm{(eff)}}}}
\def\Cpeff#1{\ensuremath{\mathcal{C}_{#1}^{'\mathrm{(eff)}}}}

\def\Opep#1{\ensuremath{\mathcal{O}_{#1}^{'}}}

\def\BdbKsmm{\ensuremath{\Bdb \to \Kstarzb\mup\mun}\xspace}
\def\BdbKsll{\ensuremath{\Bdb \to \Kstarzb\ellell}\xspace}

\def\AFB{\ensuremath{A_{\mathrm{FB}}}\xspace}
\def\AT#1{\ensuremath{A_T^{#1}}\xspace}
\def\C#1{\ensuremath{\mathcal{C}_{#1}}}
\def\Cp#1{\ensuremath{\mathcal{C}_{#1}^{'}}}
\def\Ceff#1{\ensuremath{\mathcal{C}_{#1}^{\mathrm{(eff)}}}}
\def\Cpeff#1{\ensuremath{\mathcal{C}_{#1}^{'\mathrm{(eff)}}}}

\def\Opep#1{\ensuremath{\mathcal{O}_{#1}^{'}}}

\def \all{A_{0L}}
\def \alr{A_{0R}}
\def \apl{A_{\| L}}
\def \apr{A_{\| R}}
\def \appl{A_{\bot L}}
\def \appr{A_{\bot R}}
\def \al{A_0}
\def \ap{A_{\|}}
\def \app{{A}_{\bot}}

\def \AT#1{\ensuremath{A_T^{\left(#1\right)}}\xspace}

\def \be{\begin{equation}}
\def \ee{\end{equation}}
\def \bea{\begin{eqnarray}}
\def \eea{\end{eqnarray}}
\def \ben{\begin{enumerate}}
\def \een{\end{enumerate}}
\def \bit{\begin{itemize}}
\def \eit{\end{itemize}}
\def \Bbar{\overline{\kern -0.24em B}}

\def \all{A_{0L}}
\def \alr{A_{0R}}
\def \apl{A_{\| L}}
\def \apr{A_{\| R}}
\def \appl{A_{\bot L}}
\def \appr{A_{\bot R}}
\def \al{A_0}
\def \ap{A_{\|}}
\def \app{{A}_{\bot}}
\def \be{\begin{equation}}
\def \ee{\end{equation}}
\def \bea{\begin{eqnarray}}
\def \eea{\end{eqnarray}}
\def \ben{\begin{enumerate}}
\def \een{\end{enumerate}}
\def \bit{\begin{itemize}}
\def \eit{\end{itemize}}
\def \Bbar{\overline{\kern -0.24em B}}

\def \all{A_{0L}}
\def \alr{A_{0R}}
\def \apl{A_{\| L}}
\def \apr{A_{\| R}}
\def \appl{A_{\bot L}}
\def \appr{A_{\bot R}}
\def \al{A_0}
\def \ap{A_{\|}}
\def \app{{A}_{\bot}}
\def \be{\begin{equation}}
\def \ee{\end{equation}}
\def \bea{\begin{eqnarray}}
\def \eea{\end{eqnarray}}
\def \ben{\begin{enumerate}}
\def \een{\end{enumerate}}
\def \bit{\begin{itemize}}
\def \eit{\end{itemize}}
\def \Bbar{\overline{\kern -0.24em B}}

\def \all{A_{0L}}
\def \alr{A_{0R}}
\def \apl{A_{\| L}}
\def \apr{A_{\| R}}
\def \appl{A_{\bot L}}
\def \appr{A_{\bot R}}
\def \al{A_0}
\def \ap{A_{\|}}
\def \app{{A}_{\bot}}
\def \be{\begin{equation}}
\def \ee{\end{equation}}
\def \bea{\begin{eqnarray}}
\def \eea{\end{eqnarray}}
\def \ben{\begin{enumerate}}
\def \een{\end{enumerate}}
\def \bit{\begin{itemize}}
\def \eit{\end{itemize}}
\def \Bbar{\overline{\kern -0.24em B}}

\def \all{A_{0L}}
\def \alr{A_{0R}}
\def \apl{A_{\| L}}
\def \apr{A_{\| R}}
\def \appl{A_{\bot L}}
\def \appr{A_{\bot R}}
\def \al{A_0}
\def \ap{A_{\|}}
\def \app{{A}_{\bot}}
\def\euro#1#2#3{{Eur. Phys. J. C} {\bf #1}, #3 (#2)}
\def\eurodirect#1#2#3{{EPJdirect} {\bf C#1}, #3 (#2)}

\def\ibid#1#2#3{{\it ibid.\/}~{\bf#1}, #3 (#2)}
\def\ib#1#2#3{{\bf#1}, #3 (#2)}

\def\pl#1#2#3{{Phys.~Lett. B}~{\bf #1}, #3 (#2)}

\def\prd#1#2#3{{Phys.~Rev. D}~{\bf #1}, #3 (#2)}

\newcommand{\AmS}{{\protect\the\textfont2
  A\kern-.1667em\lower.5ex\hbox{M}\kern-.125emS}}

\title{New physics search in the LHCb era \hspace{3cm} {\tiny CERN-PH-TH/2008-174, SLAC-PUB-13407}}

\author{Tobias Hurth\address{CERN, Theory Group, CH-1211 Geneva 23, Switzerland\\ SLAC, Stanford University, Stanford, CA 94309, USA}%
                            }
       
\begin{document}

\begin{abstract}
We present theoretical and experimental preparations for an indirect search for new physics 
(NP) using the rare decay~\BdbKsmm. We design new observables  
 with  very small theoretical uncertainties and good experimental
resolution. 
\vspace{1pc}
\end{abstract}

\maketitle

\section{Introduction}

At the start of the LHC we are confronted with the experimental fact that 
all data on flavour observables from Babar, Belle, CLEO  and also from 
D0 and CDF are consistent with the Standard Model (SM) predictions \cite{Buchalla:2008jp}.
This implies that generic new physics  (NP) contributions  
in $K -\bar K $ mixing for example guide us to a new-physics
scale of $10^{3} - 10^{4}$ TeV  depending if the new contributions enter at 
loop-  or tree-level.
This is in strong contrast to the working hypothesis of the LHC that there
is NP  "around the corner" at $1$ TeV  in order to stabilise the Higgs 
boson mass. 
Therefore, any NP at the $1$ TeV scale has to have a non-generic
flavour structure and we have to understand why new flavour-changing neutral
currents (FCNC) are suppressed. 
Rare decays and CP violating observables allow an analysis of this flavour
problem.

The  crucial problem in the new physics search within flavour physics is 
the optimal separation of NP  effects from hadronic uncertainties.
It is well known that inclusive decay modes are dominated by partonic 
contributions; non-perturbative corrections are in general 
rather small~\cite{Hurth:2003vb,Hurth:2007xa}. Also ratios of exclusive  
decay  modes such as 
asymmetries are well suited for the new-physics search. Here large 
parts of the hadronic  uncertainties partially cancel out; for example, 
there are CP asymmetries  that are governed by one weak phase only; thus 
the hadronic matrix  elements cancel out completely. 
It is the latter opportunity  which represents the general strategy followed by \lhcb 
for the construction of 
theoretically clean observables.

In this letter we briefly discuss the theoretical and experimental preparations for an
indirect NP search using the rare decay~\BdbKsmm based on the QCDf/SCET approach
\cite{Egede:2008uy}.  QCD corrections are included at the next-to-leading order level
and also the impact of the unknown $\Lambda/m_b$ corrections is  made explicit.

The  exclusive decay~\BdbKsmm  was  first observed at Belle \cite{Ishikawa:2003cp}.  
It offers a rich phenomenology of various kinematic distributions beyond the 
measurement of the branching ratio.  We note that some \mbox{experimental} analyses of 
those angular  distributions are already presented by the $B$ factories
\cite{Aubert:2003cm,Ishikawa:2006fh,Aubert:2006vb,Aubert:2008ju}.  Those experimental results already have a significant impact on the
model-independent constraints within the minimal flavour violation 
approach \cite{Hurth:2008jc}.

Large increase in statistics at  \lhcb~\cite{Dickens:2007ka,Dickens:2007zz,Egede:2007zz} for \BdbKsmm will make  much higher precision measurements possible. There are also great
opportunities at the future (Super-)$B$ factories in this
respect~\cite{Browder:2007gg,Bona:2007qt,Hewett:2004tv,Akeroyd:2004mj}.

Previously proposed angular distributions and \CP violating observables in
\BdbKsmm are reviewed in Ref.~\cite{SuperbKrueger}, and more recently QCDf
analyses of such angular distributions \cite{Kruger:2005ep,Lunghi:2006hc} and
CP violating observables~\cite{Bobeth:2008ij},  based on the NLO results 
in Ref.  \cite{Beneke:2001at}, were presented.

\section{QCD factorization, SCET}

Regarding the hadronic matrix elements of exclusive modes,  
the method of QCD-improved \mbox{factorization} (QCDf) has been 
systemized for non-leptonic decays in the heavy-quark limit. This method 
allows for a perturbative calculation of QCD corrections to naive 
factorization and is the basis for the up-to-date predictions for exclusive 
rare $B$ decays in general \cite{Beneke:1999br}.  

A  quantum 
field theoretical framework was proposed 
-- known under the name of
soft-collinear effective field theory (SCET) -- which allows for a deeper 
understanding of the QCDf  approach \cite{SCETa,SCETb}.
 In contrast to the  
heavy-quark effective theory (HQET),  SCET does not correspond to a local operator expansion.  
 HQET is only  applicable to $B$ decays, when the energy transfer 
to light hadrons  is small, for example to  $B \rightarrow D$ transitions 
at small recoil to the $D$ meson.  HQET is not applicable, when 
 some of the outgoing, light particles have momenta of order $m_b$; 
then one faces a multi scale problem that  can be tackled within SCET.  

There  are three  scales: 
a) $\Lambda = {\rm few} \times \Lambda_{\rm QCD}$
the {\it soft} scale set by the typical energies and
momenta of the light \mbox{degrees} of freedom in the hadronic bound states;
b) $m_b$ the {\it hard}\/ scale set by the heavy-$b$\/-quark mass and  
also by the energy of the final-state hadron in 
the $B$\/-meson rest frame;  and 
c) the hard-collinear scale $\mu_{\rm hc}=\sqrt{m_b
\Lambda}$ appears through interactions between soft and energetic
modes in the initial and final states. 
The dynamics of hard and hard-collinear
modes can be described perturbatively in the heavy-quark limit 
$m_b \to \infty$.
Thus, SCET describes $B$ decays to light hadrons with energies 
much larger than their masses, assuming that their constituents have 
momenta collinear to the hadron momentum.

However, we emphasize  that 
within the QCDf/SCET  approach, a general, quantitative method to estimate the
important $\Lambda/m_b$ corrections to the heavy-quark limit is missing
which  has important phenomenological consequences.

A  careful choice of observables needs to be made to take full advantage of the
exclusive decay 
\BdbKsmm , 
as only in certain ratios such as \CP and forward-backward
asymmetries, the hadronic uncertainties cancel out making such ratios the only observables that are highly sensitive to NP. 

Within the QCDf/SCET approach one finds
crucial form factor relations 
\cite{large:energy:limit}
which simplify the
theoretical structure of various kinematical distributions such that,  at least
at the leading order (LO) level any hadronic uncertainties cancel out. A
well-known example of this is the zero-crossing of the forward-backward
asymmetry.  In \cite{Egede:2008uy}  new observables of this kind in the \BdbKsmm decay 
were proposed which have
very small theoretical uncertainties and good experimental
resolution. The only difference to the forward-backward asymmetry is that within these 
new observables the hadronic form factors cancel out for {\it all} values of the dilepton mass.

\section{Theoretical preliminaries}

The decay \BdbKsll with $\Kstarzb \to \Km \pip$ on the mass shell  is
completely described by four independent kinematic variables, the lepton-pair
invariant mass squared, \qsq, and the three angles $\theta_l$, $\theta_{K}$,
$\phi$. Summing over the spins of the final particles, the differential decay
distribution of \BdbKsll can be written as \cite{FK:etal,dmitri,CS:etal,4body:mass}:
\begin{eqnarray}
  \label{diff:four-fold}  \nonumber
  \frac{d^4\Gamma_{\Bdb}}{dq^2\,d\theta_l\, d\theta_K\, d\phi} = 
  \frac{9}{32 \pi} I(q^2, \theta_l, \theta_K, \phi) \sin\theta_l\sin\theta_K  \nonumber
\end{eqnarray}
with 
\begin{eqnarray}
\label{funcs:i}
I &=& I_1 + I_2\cos 2\theta_l + I_3 \sin^2\theta_l\cos 2\phi \nonumber \\  & &+  I_4 \sin 2\theta_l \cos\phi + I_5 \sin\theta_l\cos\phi \nonumber \\& &
+ I_6 \cos\theta _l  +  I_7 \sin\theta_l\sin\phi \nonumber \\&&+ I_8 \sin 2\theta_l \sin\phi + I_9 \sin^2\theta_l\sin 2\phi. 
\end{eqnarray}
The $I_i$ depend on products of the seven complex $K^*$ spin amplitudes,
$A_{\bot L/R}$, $A_{\| L/R}$, $A_{0 L/R}$, $A_t$ with each of these a function of \qsq;  
the explicit formulae are given in the appendix.
 $A_t$ is related to the time-like component of the virtual $K^*$, which does not
contribute in the case of massless leptons and can be neglected if the lepton
mass is small in comparison to the mass of the lepton pair.  We will consider
this case in our present \mbox{analysis}.

The six complex $K^*$ spin amplitudes 
of the  massless case are related to the well-known helicity amplitudes 
(used for example in \cite{dmitri,CS:etal,ali:safir}):
\be\label{hel:trans} 
A_{\bot,\|} = (H_{+1}\mp H_{-1})/\sqrt{2}, \quad A_0=H_0. 
\ee

The crucial theoretical input we use in  our \mbox{analysis} is 
the observation that  in the limit
where the initial hadron is heavy and the final meson has a large
energy \cite{large:energy:limit} the hadronic form factors  can be 
expanded in the small ratios $\Lambda_{\mathrm{QCD}}/m_b$ and
$\Lambda_{\mathrm{QCD}}/E$, where $E$ is the energy of
the light meson. 
Neglecting  corrections of order $1/m_b$ and $\alpha_s$, 
the seven a priori independent $B\to K^*$ form factors  
 reduce to two universal form factors $\xi_{\bot}$ and  $\xi_{\|}$
\cite{large:energy:limit,Beneke:2001wa} and one finds 
that the spin  amplitudes 
at leading order in $1/m_b$ and $\alpha_s$ have a very
simple form:
\begin{eqnarray}
A_{\bot L,R}&=&  \sqrt{2} N m_B(1- \hat{s}) \times 
\bigg[(\Ceff{9}\mp\C{10})\nonumber\\  & &+ \frac{2\hat{m}_b}{\hat{s}} (\Ceff{7} + \Cpeff{7}) 
\bigg]\xi_{\bot}(E_{K^*}),\nonumber
\end{eqnarray}
\begin{eqnarray}
A_{\| L,R} &=& -\sqrt{2} N m_B (1-\hat{s})\times \bigg[(\Ceff{9}\mp \C{10}) \nonumber\\ 
& &+\frac{2\hat{m}_b}{\hat{s}}(\Ceff{7} -\Cpeff{7}) \bigg]\xi_{\bot}(E_{K^*}),
\nonumber \end{eqnarray}
\begin{eqnarray}\label{Ksternamplitudes}
A_{0L,R}&=& -\frac{Nm_B }{2 \hat{m}_{K^*} \sqrt{\hat{s}}} (1-\hat{s})^2\bigg[(\Ceff{9}\mp \C{10})   \nonumber\\ & &+2
\hat{m}_b (\Ceff{7} -\Cpeff{7}) \bigg]\xi_{\|}(E_{K^*}),
\end{eqnarray}
with $\hat{s} =  \qsq/m_B^2$, $\hat{m}_i =  m_i/m_B$. Here we neglected  
terms of $O(\hat{m}_{K^*}^2)$. 
It is important to mention that  the theoretical simplifications are restricted to the kinematic region
  in which the energy of the $K^*$ is of the order of the heavy quark mass,
  i.e.~$\qsq \ll m_B^2$. Moreover, the influences of very light resonances
  below 1\gev  question the QCD factorization results in that region.
  Thus, we will confine our analysis of all observables to the dilepton mass
  in the range  $1 {\rm GeV}^2  \leq  q^2 \leq  6  {\rm GeV}^2$.

\section{Construction of theoretically clean observables}

By inspection one finds that the distribution functions $I_i$ in the differential
decay distribution (see  Eq.~(\ref{Iq2}))  are  {\it invariant}
under three symmetry transformations which are given explicitly 
in the appendix (see Eqs.~(\ref{Sym1}-\ref{Sym3})). This implies that only 9 
of the 12 $K^*$ spin amplitudes are {\it independent} and that they 
can be fixed by an full  angular fit to the 9  independent coefficients 
of the differential decay distribution. 
Another  direct consequence is that 
 any observable based on the
differential decay distribution has also to be invariant under the same
symmetry transformations.

Besides this  mandatory criterium there are further criteria required for
an interesting observable. 
[{\bf Simplicity:}] A simple functional dependence on the 9 independent
  measurable distribution functions; at best it should depend only from one or
  two in the numerator and denominator of an asymmetry.
[{\bf Cleanliness:}] At leading order in $\Lambda/m_b$ and in $\alpha_s$ the
  observable should be independent of any form factor, at best for all
  $q^2$. Also the influence of symmetry-breaking corrections at order
  $\alpha_s$ and at order $\Lambda/m_b$ should be minimal.
[{\bf Sensitivity:}] The sensitivity to the $\Cpeff{7}$ Wilson coefficient
  representing NP with another chirality than in the SM should be maximal. 
  [{\bf Precision:}] The experimental precision obtainable should be good enough
  to distinguish different NP models.

In the limit where the \Kstarzb meson has a large energy, only two independent
form factors occur in $A_{0 L/R}$ and in $A_{\bot L/R}$ and $A_{\|
  L/R}$. Clearly, any ratio of two of the nine measurable distribution functions
proportional to the same form factor fulfil the criterium of symmetry,
simplicity, and theoretical cleanliness up to $\Lambda/m_b$  and $\alpha_s$ 
corrections. However, the third criterium, a sensitivity to a special kind of
NP and the subsequent requirement of experimental precision, singles
out particular combinations. In \cite{Egede:2008uy}
we focused on new right-handed
currents. Other NP sensitivities may single out other observables as
will be analysed in a forthcoming paper~\cite{secondpaper}.

\section{Results}

The first surprising result  is that the previously proposed quantity 
$\AT{1}$~\cite{dmitri}, \begin{equation}
    \label{eq:AT1Def}
    \AT{1}=\frac{\Gamma_{-}-\Gamma_{+}}{\Gamma_{-}+\Gamma_{+}}
    =\frac{-2\Re(\ap^{}\app^*)}{|\app|^2 + |\ap|^2} \ .
  \end{equation}
  with $\Gamma_\pm = |H^L_{\pm1}|^2 +
  |H^R_{\pm1}|^2$  
does not fulfil the most important
  criterium of symmetry while it has very attractive new physics sensitivity
  \cite{Kruger:2005ep,Lunghi:2006hc}.
 Therefore,  it is not possible to extract
$\AT{1}$  from the full angular distribution which is constructed after summing 
over the spins of the final particles. 
Because it seems practically not possible to measure the helicity of the final states
on a \textit{event-by-event} basis, 
$\AT{1}$ cannot  be measured at either \lhcb or at a Super-$B$ factory with electrons or 
muons in the final state.

\begin{figure}
\begin{center}
\includegraphics[width=0.39\textwidth]{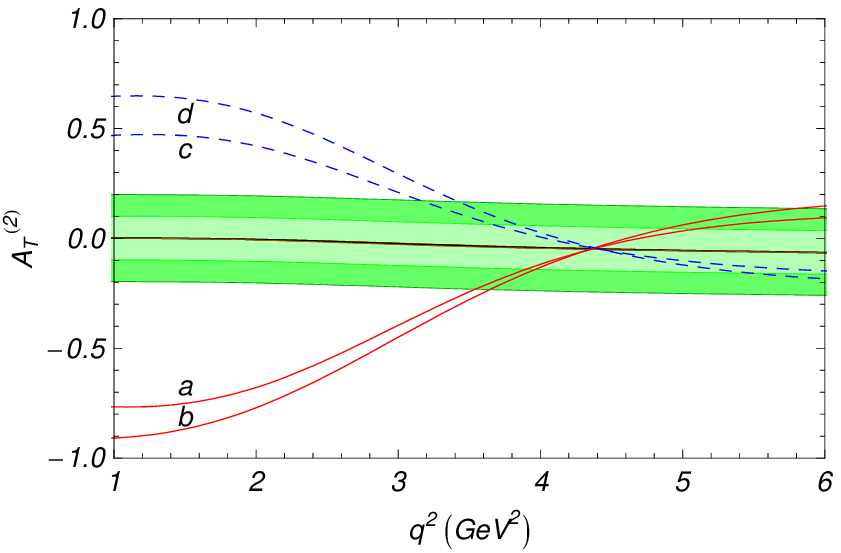}
\includegraphics[width=0.39\textwidth]{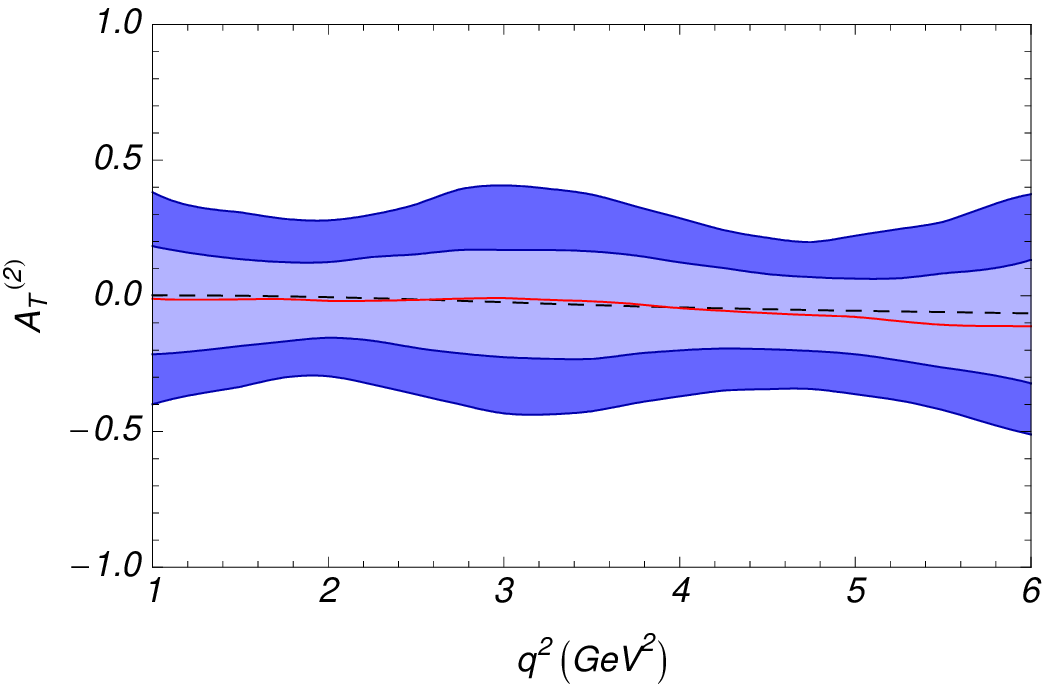}
\end{center}
\vskip -0.5cm \caption{For \AT{2}, theoretical errors (top), experimental errors (bottom) as a function of the squared dimuon  mass, see text for details. }
  \label{fig:AT2}
\end{figure}

\begin{figure}
\begin{center}
\includegraphics[width=0.39\textwidth]{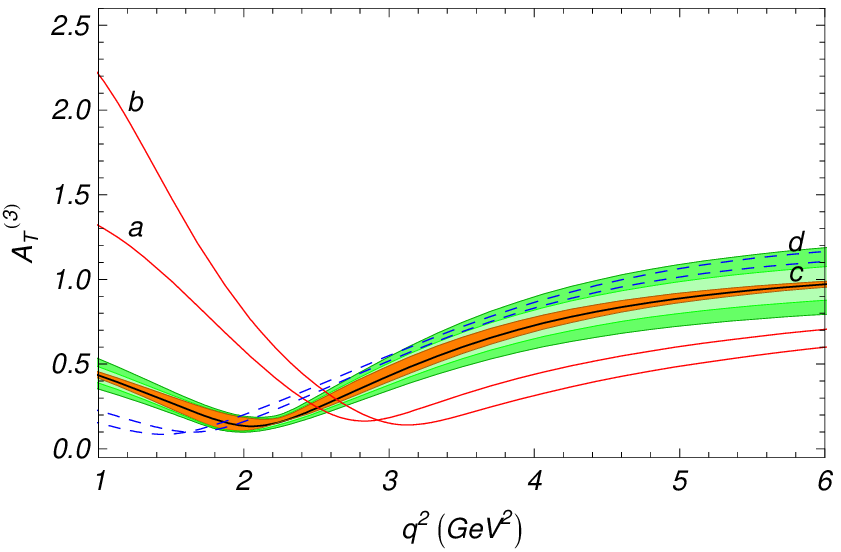}
\includegraphics[width=0.39\textwidth]{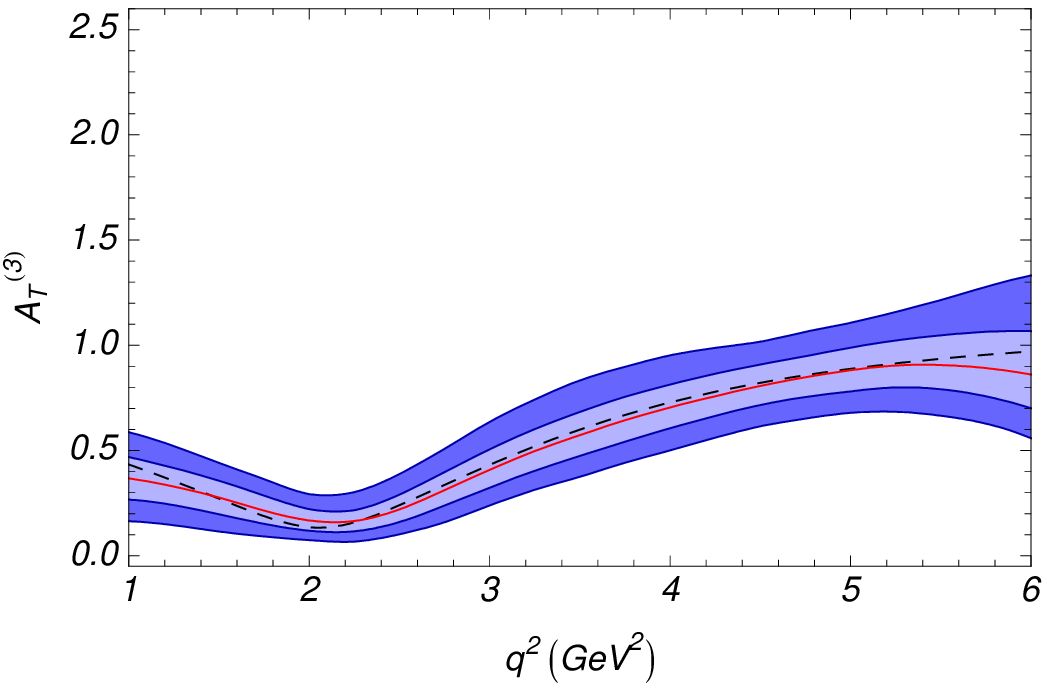}
\end{center}
\vskip -0.5cm \caption{\AT{3}, as in Fig.1.}
\label{fig:AT3}
\end{figure}

\begin{figure}
\begin{center}
\includegraphics[width=0.39\textwidth]{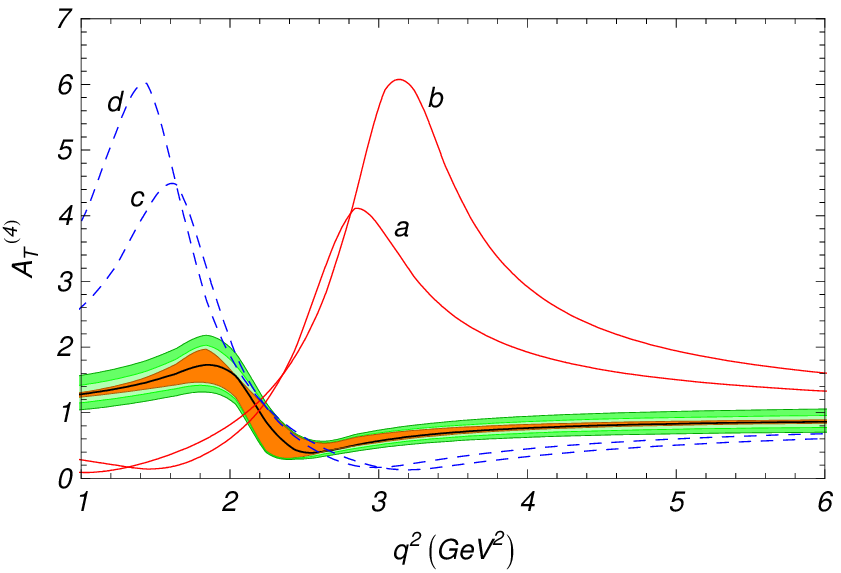}
\includegraphics[width=0.39\textwidth]{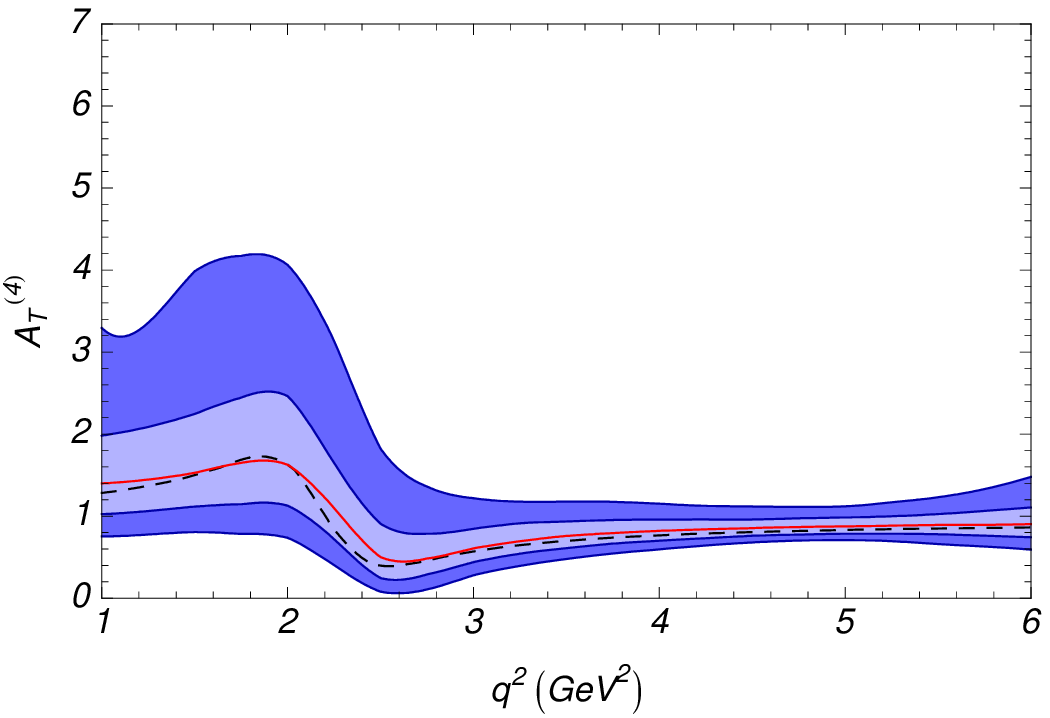}
\end{center}
\vskip -0.5cm \caption{ \AT{4}, as in Fig.1.}                   
\label{fig:AT4}
\end{figure}

\begin{figure}
\begin{center}
\includegraphics[width=0.39\textwidth]{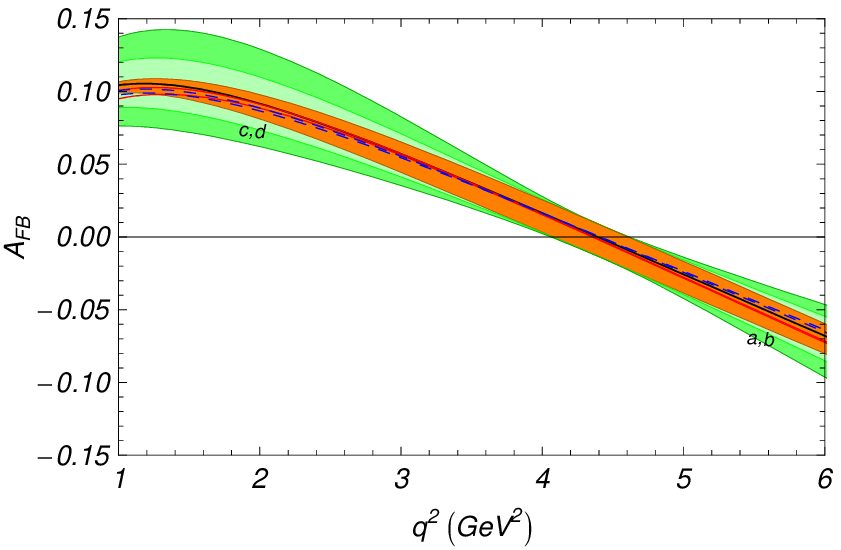}
\includegraphics[width=0.39\textwidth]{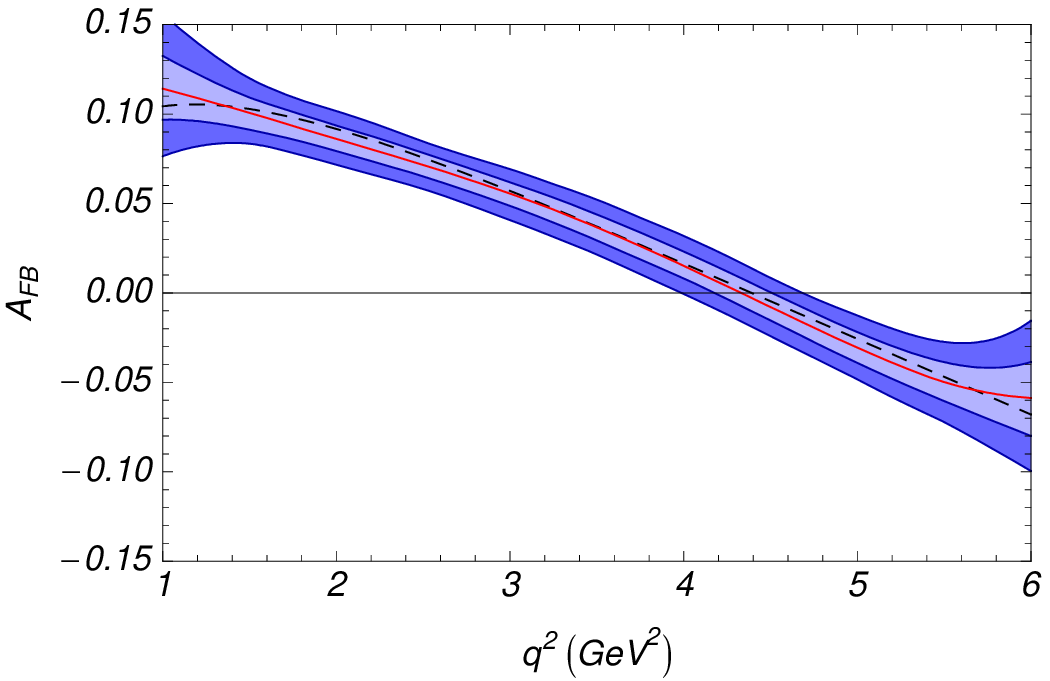}
\end{center}
\vskip -0.5cm \caption{$A_{FB}$,  as in Fig.1.}     
  \label{fig:Afb}
\end{figure}

\begin{figure}
\begin{center}
\includegraphics[width=0.39\textwidth]{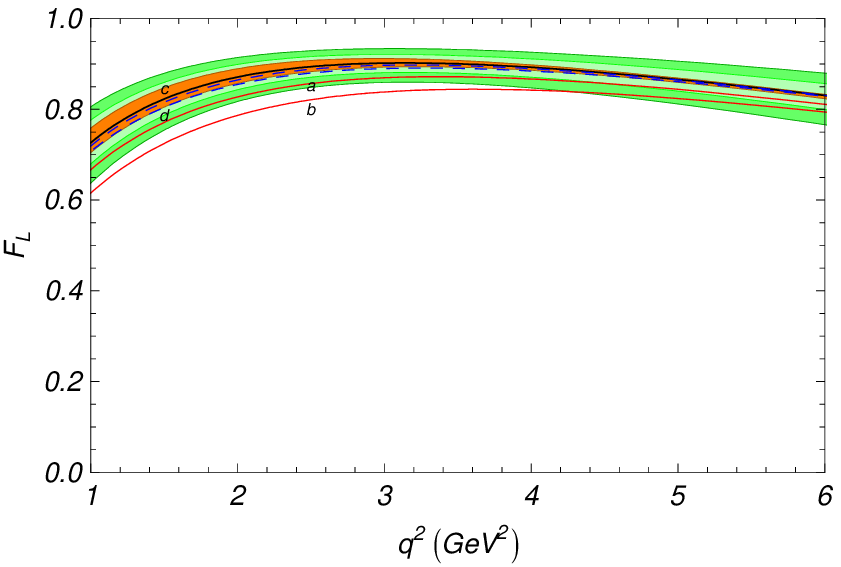}
\includegraphics[width=0.39\textwidth]{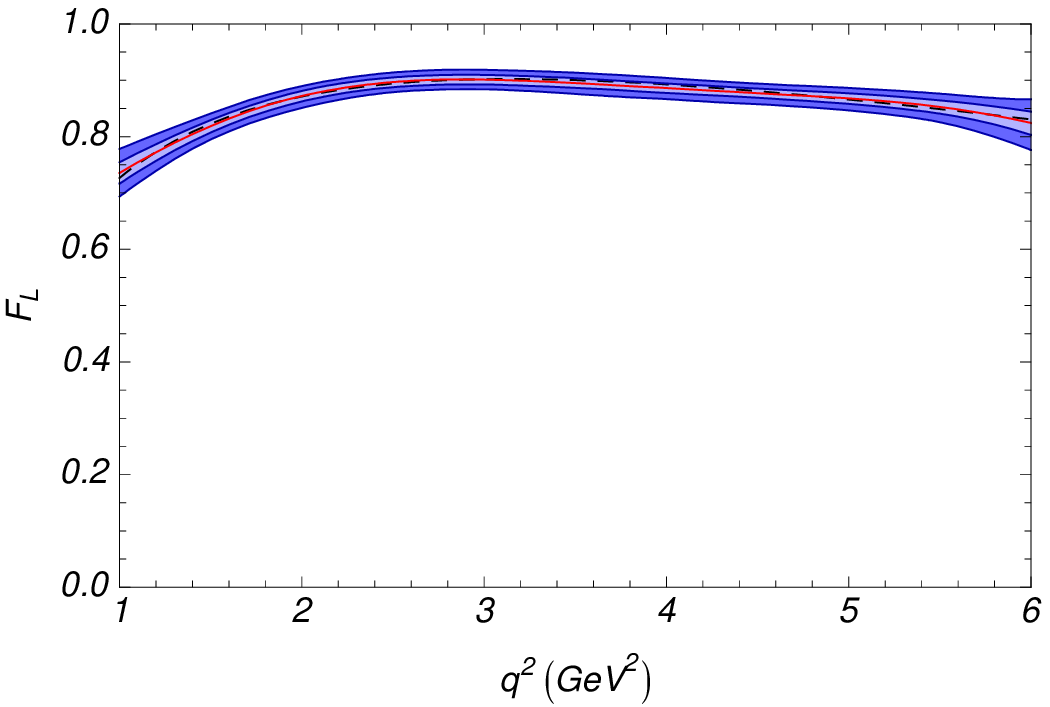}
\end{center}
\vskip -0.5cm \caption{$F_L$,  as in Fig.1.}
\label{fig:FL}
\end{figure}

One finds that the well-known quantities, the forward-backward asymmetry $A_{\rm FB}$ and the \mbox{longitudinal} $K^*$ polarization $F_L$ fulfill the symmetry but they  include larger 
\mbox{theoretical} uncertainties due to the fact that the form factors do not cancel at leading order 
level for all dilepton masses. Moreover, the sensitivity to right-handed currents is marginal  
as it is shown below,
\begin{equation}
    \label{eq:AFBdef}
    \AFB = \frac{3}{2}\frac{\Re(\apl\appl^*) - \Re(\apr\appr^*)}
    {|\al|^2 + |\ap|^2 + |\app|^2}       
  \end{equation}
  where  for $i,j  = 0, \|, \perp$ 
  \begin{eqnarray}
    A_i A^*_j\equiv A^{}_{i L}(q^2) A^*_{jL}(q^2)+ A^{}_{iR}(q^2) A^*_{jR}(q^2)  ,\nonumber  
  \end{eqnarray}
\begin{equation}\label{def:frac:pol}
    F_L(\qsq) =  \frac{|{A}_0|^2}{|{ A}_0|^2 + |{A}_{\|}|^2
      + |A_\perp|^2} .   \end{equation}

In contrast, the following three observables,
\begin{equation}
    \label{eq:AT2Def}
    \AT{2} =\frac{|\app|^2 - |\ap|^2}{|\app|^2 + |\ap|^2} ,
  \end{equation}
\begin{equation}
  \label{eq:AT3Def}
  \AT{3} =
  \frac{|\all\apl^* + \alr^*\apr|}{\sqrt{ |\al|^2  |\app|^2}} \ ,
\end{equation}
\begin{equation}
  \label{eq:AT4Def}
  \AT{4} = \frac{|A_{0L} A_{\perp L}^* - A_{0R}^* A_{\perp R}| }{|
    A_{0L}^* A_{\|L}+A_{0R} A_{\|R}^*|} \ ,
\end{equation}
are theoretically clean for {\it all}  dilepton  masses and also
show a very high sensitivity to right-handed currents.

In the following figures the  results on  the observables, $F_L$, $A_{FB}$, 
\AT{2}, \AT{3},  and \AT{4} are illustrated:
For all the observables the theoretical sensitivity is plotted on the top 
of each \mbox{figure.} 
  The thin dark line is the central NLO result for the SM and the narrow inner
  dark (orange) band that surrounds it corresponds to the NLO SM uncertainties
  due to both input parameters and perturbative scale dependence. Light grey
  (green) bands are the estimated $\Lambda/m_b \pm 5\%$ corrections for each
  spin amplitude  while darker grey (green)
  ones are the more conservative $\Lambda/m_b \pm 10\%$ corrections. The
  curves labelled (a)--(d) correspond to four different benchmark points
  in the MSSM for righthanded currents (for more details see \cite{Egede:2008uy}).
 The experimental sensitivity for a dataset corresponding to 10\invfb of
  \lhcb data is given in each figure on the  bottom, assuming the SM. Here the solid
  (red) line shows the median extracted from the fit to the ensemble of data
  and the dashed (black) line shows the theoretical input distribution.  The
  inner and outer bands correspond to 1$\sigma$ and 2$\sigma$ experimental
  errors.

The observables 
\AT{3} and \AT{4}  offer sensitivity to the longitudinal
spin amplitude $A_{0 L,R}$ in a controlled way compared to the old observable
$F_L$: the dependence on both the parallel and perpendicular 
soft form factors $\xi_\|(0)$ and $\xi_\perp(0)$ cancels at LO. A residual of this
dependence may appear at NLO, but as shown in Figs.~\ref{fig:AT3} and \ref{fig:AT4},  
it is basically negligible. 
It is also remarkable that for \AT{3} and \AT{4} at low \qsq the
impact of this uncertainty is less important than the 
uncertainties due to input parameters and scale dependence.
The observables \AT{3} and \AT{4} also present a different sensitivity to
\Cp{7} via their dependence on  $A_{0 L,R}$ compared with \AT{2}.  
This may allow for a particularly interesting cross check of the sensitivity 
to this chirality flipped operator \Opep{7}; for instance, new contributions 
coming from tensor scalars and  pseudo-scalars will behave differently 
among the set of observables.

Another remarkable point that becomes clear when comparing the set of clean
observables \AT{2}, \AT{3} and \AT{4} versus the old observables $F_L$
and $A_{FB}$ concerns the potential discovery of NP, in particular of new right-handed currents.  
There are large deviations from the SM curve
from the ones of the four supersymmetric benchmark points. A large deviation 
from the SM for \AT{2}, \AT{3} or \AT{4} can thus show the presence of right-handed currents in a way that is not  possible with $F_L$ or $A_{FB}$.  In the latter cases the deviations from the
SM prediction of the same four representative curves are marginal.

In the experimental plots we find a good agreement between the central values
extracted from the fits and the theoretical input. Any deviations seen are
small compared to the statistical uncertainties.  The experimental resolution for $F_L$ is
very good but with the small deviations from the SM expected this is not
helpful in the discovery of new right-handed currents. 
Comparing the theoretical and experimental figures for the other 
observables it can be seen that in
particular \AT{3} show great promise to distinguish between NP
models.

\begin{figure}
\includegraphics[width=0.45\textwidth]{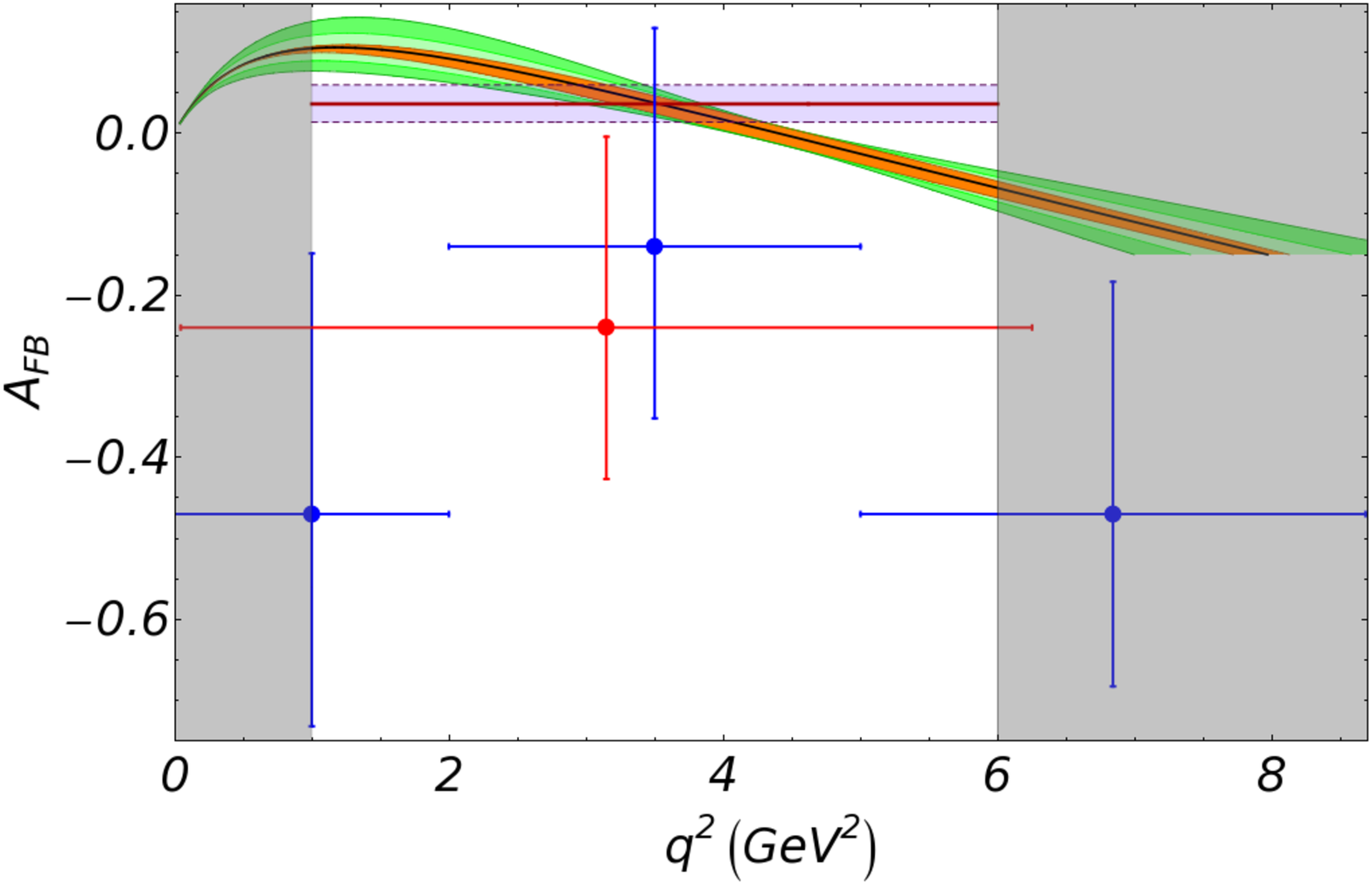}
\includegraphics[width=0.45\textwidth]{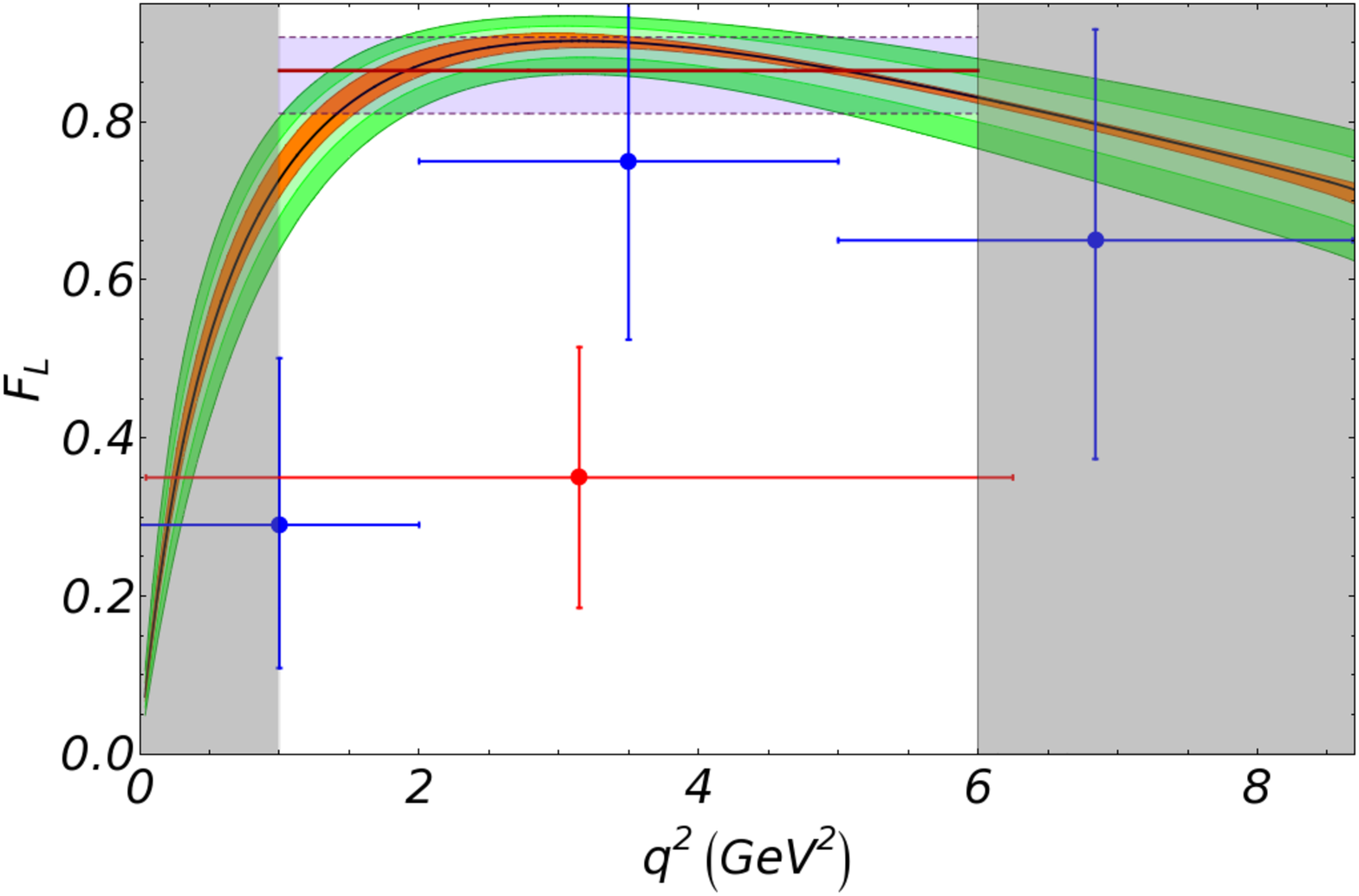}
\vskip -0.5cm \caption{Belle (black/blue) and BaBar (grey/red) data points on $F_L$ and on $A_{FB}$ with SM predictions and weighted SM averages over the bin $q^2 \in [1\gevgev, 6\gevgev]$}
\label{fig:bellebabar}
\end{figure}

Finally, let us  mention that the old observables
$F_L$ and $A_{FB}$ are already accessible to the 
BaBar\cite{Aubert:2008ju,Flood} and Belle\cite{Wei} experiments.  The first  
measurements  are shown  
in Fig.~\ref{fig:bellebabar} with the SM predictions and the 
weighted SM averages over the bin $q^2 \in [1\gevgev, 6\gevgev]$. 
All the present data  is compatible with the SM predictions. 
For example, the first measurement  of the Babar collaboration on $F_L$ 
in the low-$q^2$ region is given  as an average over the bin $q^2 \in [4m_\mu^2,6.25\gevgev]$:
\begin{equation}
F_L ([4m_\mu^2, 6.25\gevgev]) = 0.35 \pm 0.16\pm 0.04; 
\end{equation}

while the theoretical  average, weighted over the rate, 
using  the bin, $q^2 \in [1\gevgev, 6\gevgev]$, based on our results is given by: 
\begin{equation}
F_L ([1\gevgev, 6\gevgev]) = 0.86^{+0.04}_{-0.05}. 
\end{equation}
Here, one should keep in mind that the spectrum below 1\gevgev is 
theoretically problematic due to the influence of very light resonances; moreover the rate and also the polarisation $F_L$ 
are changing dramatically around 1\gevgev. 
Therefore, we strongly recommend to 
use the standard bin from 1\gevgev to 6\gevgev in all future measurements.

\section{Summary} 
\label{sec:summary}
The full angular analysis of the decay \BdbKsmm at the \lhcb experiment 
offers great opportunities for the 
new physics search.  New observables can be designed to be sensitive to a specific 
kind of NP operator within the model-independent analysis using the effective field theory approach.
The new observables  \AT{2}, \AT{3} and \AT{4}  are shown to be highly sensitive to right handed
currents.  Clearly,  theoretical progress on the $\Lambda/m_b$ corrections would enhance their
sensitivity significantly and would be highly desirable in view of a possible  upgrade of the \lhcb experiment. Moreover,we have shown that the previously discussed angular distribution
\AT{1} cannot be measured at either \lhcb or at a Super-$B$ factory.

\section*{Acknowledgement}
I   thank Ulrik Egede, Joaquim Matias,  Marc \mbox{Ramon}, and Will 
Reece for a pleasant collaboration. Support of the European network 
Heptools is gratefully acknowledged. 

\section*{Appendix}

We add here the explicit formula for the distribution functions and their symmetries: 

  In the massless limit, the distribution functions $I_i$ depend on products of the six complex $K^*$ spin amplitudes, $A_{\bot L/R}$, $A_{\| L/R}$, $A_{0 L/R}$:
 \begin{eqnarray} \label{Iq2}
    I_1 &=&  \frac{3}{4}\left(|\appl|^2 + |\apl|^2 + (L\to R)\right)
    \sin^2\theta_K +\nonumber\\
    & &+ \left(|\all|^2 +|\alr|^2 \right) 
    \cos^2\theta_K \nonumber\\ 
    &\equiv& a \sin^2\theta_K+ b \cos^2\theta_K, \nonumber \\
    I_2 &=&
    \frac{1}{4}( |\appl|^2+
    |\apl|^2)\sin^2\theta_K+\nonumber\\    &&- |\all|^2\cos^2\theta_K + (L\to
    R)\nonumber \\ 
    &\equiv& c \sin^2\theta_K+ d \cos^2\theta_K , \nonumber\\
    I_3 & = &\frac{1}{2}\bigg[(|\appl|^2 - |\apl|^2 )\sin^2\theta_K +
    (L\to R)\bigg]\nonumber\\   &\equiv& e \sin^2\theta_K,\nonumber\\
    I_4 & = &
    \frac{1}{\sqrt{2}}\bigg[\Re
    (\all^{}\apl^*) \sin 2\theta_K + (L\to R)\bigg] \nonumber \\  &\equiv& f \sin
    2\theta_K,  \nonumber\\
    I_5 & = &\sqrt{2}\bigg[\Re(\all^{}\appl^*) \sin2\theta_K -
    (L\to R)\bigg]\nonumber\\ &\equiv& g \sin 2\theta_K,\nonumber \\
    I_6 & = &
    2\bigg[\Re
    (\apl^{}\appl^*) \sin^2\theta_K - (L\to R)\bigg]\nonumber\\ &\equiv&  h
    \sin^2\theta_K, \nonumber\\
    I_7 & = & \sqrt{2} \bigg[\Im (\all^{}\apl^*) \sin2\theta_K -
    (L\to R)\bigg]\nonumber\\\  &\equiv&  j \sin 2\theta_K, \nonumber\\
    I_8 & = &
    \frac{1}{\sqrt{2}}\bigg[\Im
    (\all^{}\appl^*) \sin2\theta_K + (L\to R)\bigg]\nonumber\\ &\equiv& k \sin
    2\theta_K, \nonumber\\
    I_9 & = & \bigg[\Im (\apl^{*}\appl) \sin^2\theta_K +
    (L\to R)\bigg] \nonumber\\   & \equiv&  m \sin^2\theta_K. 
  \end{eqnarray}
Taking into account $a=3c$
and $b=-d$, we are left with 9  independent parameters which can be fixed
experimentally in a full angular fit. 

The distribution functions are
{\it invariant} under the following three independent symmetry transformations
of the spin amplitudes as one easily verifies,  using the explicit formulae given above: (1) a global phase transformation of the $L$-amplitudes
\begin{eqnarray}
  \label{Sym1}  
  A^{'}_{\bot L} & = & e^{i \phi_L} A_{\bot L},\nonumber  \\
  A^{'}_{\| L} & = & e^{i \phi_L} A_{\| L},  \nonumber \\   A^{'}_{0  L} &=&
  e^{i \phi_L} A_{0  L};
\end{eqnarray}
(2) a global  transformation of the $R$-amplitudes
\begin{eqnarray}
  \label{Sym2}  
  A^{'}_{\bot R} &=& e^{i \phi_R} A_{\bot R},\nonumber \\  A^{'}_{\| R} &=& e^{i \phi_R} A_{\| R}, \nonumber\\   A^{'}_{0  R} &=& e^{i \phi_R} A_{0  R};
\end{eqnarray}
and (3) a continuous $L \leftrightarrow R$ rotation
  \begin{eqnarray}
\label{Sym3}
    A^{'}_{\bot L} &=& + \cos\theta A_{\bot L}  + \sin\theta  A^*_{\bot R}\nonumber \\
    A^{'}_{\bot R} &=& -  \sin\theta A^*_{\bot L}  +  \cos\theta  A_{\bot R}\nonumber \\
    A^{'}_{0 L} &=&  + \cos\theta A_{0 L}  - \sin\theta  A^*_{0 R}\nonumber \\
    A^{'}_{0 R} &=& + \sin\theta A^*_{0 L}  +  \cos\theta  A_{0 R}\nonumber \\
    A^{'}_{\| L} &=& + \cos\theta A_{\| L}  - \sin\theta  A^*_{\| R}\nonumber \\
    A^{'}_{\| R} &=& + \sin\theta A^*_{\| L}  +  \cos\theta  A_{\| R}.
  \end{eqnarray}

  \end{document}